\documentclass[letterpaper,twocolumn,10pt,hyphens]{article}
\usepackage{usenix}

\usepackage{hyperref}

\usepackage{xcolor}
\definecolor{backcolour}{rgb}{0.95,0.95,0.95}

\usepackage{colortbl}

\usepackage[justification=centering]{caption}

\usepackage{enumitem}

\usepackage[flushleft]{threeparttable}

\usepackage{listings}
\usepackage{subcaption}

\usepackage{xspace} 

\usepackage[ruled, linesnumbered]{algorithm2e}
\SetKwComment{Comment}{\footnotesize $\triangleright$\ }{} 
\usepackage{balance}

\definecolor{cb1}{HTML}{4D6D9D}
\definecolor{cb2}{HTML}{5D88C1}
\definecolor{cb3}{HTML}{73A7CF}
\definecolor{cb4}{HTML}{DCD5D1}
\definecolor{cb5}{HTML}{E2DFE0}
\definecolor{cb6}{HTML}{E2E5EC}

\definecolor{contrastcol1}{HTML}{c38888}
\definecolor{contrastcol2}{HTML}{b19886}
\definecolor{contrastcol3}{HTML}{f2b56b}
\definecolor{contrastcol4}{HTML}{97bc62}
\definecolor{contrastcol5}{HTML}{9bc4e2}
\definecolor{contrastcol6}{HTML}{d7d2cb}

\definecolor{links}{HTML}{000000}

\begin{document}

\date{}

\author{
{\rm Pallavi Sivakumaran}\\
Information Security Group\\
Royal Holloway, University of London\\
Email: pallavi.sivakumaran.2012@rhul.ac.uk
\and
{\rm Jorge Blasco}\\
Information Security Group\\
Royal Holloway, University of London\\
Email: jorge.blascoalis@rhul.ac.uk
} 

\title{\texttt{argXtract}: Deriving IoT Security Configurations via \\Automated Static Analysis of Stripped ARM Binaries}

\maketitle

\begin{abstract}
Recent high-profile attacks on the Internet of Things (IoT) have brought to the forefront the vulnerability of ``smart'' devices, and have resulted in IoT technologies and devices being subjected to numerous security analyses. 
Many of the attacks had weak device configuration as the root cause, making the configuration of IoT devices a vector of interest for security analysis.
One potential source of rich and definitive information about the configuration of an IoT device is the device's firmware itself. 
However, firmware analysis is complex and automated firmware analyses have thus far been confined to IoT hub or gateway devices, or peripheral devices with more traditional operating systems such as Linux or VxWorks. 
Most IoT peripherals, by their very nature of being resource-constrained, lacking traditional operating systems and implementing a wide variety of communication technologies, have only been the subject of smaller-scale analyses, typically confined to a certain class or brand of device. 
Analysing peripheral firmware is further complicated by the fact that peripheral firmware files are predominantly available as \textit{stripped binaries}, without the ELF headers and symbol tables that would simplify reverse engineering. 

In this paper, we present \texttt{argXtract}, an open-source automated static analysis tool, which extracts security-relevant configuration information from stripped IoT peripheral firmware. Specifically, we
target binaries that implement the ARM Cortex-M architecture, due to its growing popularity among IoT peripherals. 
\texttt{argXtract} overcomes the challenges associated with stripped Cortex-M analysis and is able to retrieve and process arguments to security-relevant supervisor and function calls, enabling automated bulk analysis of firmware files. 
We demonstrate this via three real-world case studies. The largest case study covers a dataset of 243 Bluetooth Low Energy binaries targeting Nordic Semiconductor chipsets, while the other two focus on Nordic ANT binaries and STMicroelectronics BlueNRG binaries. The results from all three case studies reveal widespread lack of security and privacy controls in IoT, such as minimal or no protection for data, fixed passkeys and trackable device addresses. 
\end{abstract}

\section{Introduction}
\label{section:introduction}
The Internet of Things (IoT) is growing at a rapid pace, with an estimated 22 billion connected IoT devices in use around the world at the end of 2018, and projected to grow to 50 billion by 2030~\cite{Statista:2019:IoT}. 
These devices are increasingly handling user PII\footnote{Personally Identifiable Information.} and health information, and performing security-related functions. 
It is therefore imperative to fully understand the security and privacy implications of IoT deployments. 

Recent years have shown that there is legitimate cause for concern, as numerous flaws have been uncovered in IoT devices, some of which have been exploited at a large-scale (e.g., Mirai~\cite{NJCCIC:2016:Mirai, Krebs:2016:Mirai,  Chirgwin:2016:ColdMirai} and Brickerbot~\cite{Radware:2017:Brickerbot}). 
Severe vulnerabilities have also been discovered in certain cardiac devices~\cite{Larson:2017:StJudePacemaker}, baby heart monitors~\cite{Thomson:2016:BabyHeartMonitor} and webcams~\cite{Stanislav:2015:BabyCam}. 
The root cause on many occasions was poor device configuration, e.g., default passwords~\cite{antonakakis:2017:Mirai, CVE:2015:CVE-2015-2880} or poor protection for data~\cite{CVE:2019:CVE-2019-16518, CVE:2018:CVE-2018-10825, Zetter:2015:eskateboard}. 

The configuration of an IoT device can therefore be a vital source of information regarding possible vulnerabilities, and the firmware running on the device is often the most definitive source of information regarding its configuration. However, analysing IoT firmware is notoriously difficult. Firmware files are often only available as stripped binaries, i.e., files that do not contain the debugging information that simplifies analysis. Further, many IoT peripheral devices favour ARM Cortex-M processors, which contain certain features (such as inline data) that greatly complicate analysis~\cite{Jiang:2020:ArmDisassemblyTools}.

In this work, we present \texttt{argXtract}, a tool for bulk extraction of security-relevant configuration information from stripped Cortex-M binaries via a partial-knowledge automated static analysis. Concretely, \texttt{argXtract} exploits multiple vectors to determine absent information regarding the application code base and locations of inline data within a stripped binary, to overcome many of the challenges that are normally encountered when analysing stripped \mbox{Cortex-M} binaries.
It provides a generic framework for extracting arguments to a security-relevant \textit{Call of Interest} (COI). In our work COIs can be ARM supervisor calls - identified using \texttt{svc} instructions - or function calls - identified using function pattern matching. 
A \textit{template}-based approach is employed for defining argument structures, to allow for greater flexibility and extensibility. 
We stress at this point that the ability to extract configuration information in this manner should not be considered a security vulnerability. It is only if the extracted configuration itself is not secure that vulnerabilities arise.

We present a case study where we use \texttt{argXtract} to analyse the security configurations from Bluetooth Low Energy (BLE) binaries targeting Nordic Semiconductor chipsets, where configurations are made via supervisor calls. To further demonstrate the applicability of \texttt{argXtract} for other vendors and technologies, we also present smaller studies of BLE binaries targeting STMicroelectronics (ST) BlueNRG chipsets, where configurations are made via function calls, and ANT binaries targeting Nordic chipsets, with configurations via supervisor calls. These case studies were selected based on the availability of real-world binaries.

The results from our analyses reveal widespread lack of protection for BLE data, at both the link and application layers. In the few cases where link-layer protection \textit{was} present, it was via the insecure \textit{Just Works} pairing model.  The results also show that in some cases, permissions were applied inconsistently for different access mechanisms for a single data value. From a privacy standpoint, we find that over 95\% of BLE devices utilised fixed addresses, which may open up the device, and therefore the user, to the risk of tracking. The results from the ANT analysis reveal that channel encryption was not defined in any of the real-world binaries.

We summarise our main contributions as follows:
\begin{itemize}[noitemsep]
   \item We present \texttt{argXtract}, a tool for performing automated static analysis to extract configuration information from supervisor calls and function calls within stripped IoT peripheral firmware binaries (\S\ref{section:system-description}). Our tool overcomes many of the challenges unique to the analysis of stripped ARM Cortex-M binaries. We evaluate it in terms of the accuracy of function boundary estimation and correctness of results (\S\ref{section:evaluation}). 
   \item We present a concrete case study for \texttt{argXtract}: the analysis of stripped Nordic BLE binaries. We examine the results from extracting security-relevant configurations from 243 binary files, and find lack of protection for BLE data, inconsistent data access controls and serious privacy vulnerabilities (\S\ref{section:nordic}). 
   \item To demonstrate the applicability of \texttt{argXtract} for other vendors and technologies, we present additional case studies for STMicroelectronics' BlueNRG BLE platform, using function pattern matching, and for Nordic binaries targeting the ANT communication technology. These case studies also reveal poor privacy and security configurations, most significantly in one case in a user safety-related application (\S\ref{section:applicability}).
   \item We make \texttt{argXtract} available as an open-source tool for the benefit of the research community (\S\ref{section:resources}).
\end{itemize}

\section{Motivation}
\label{section:background}
Configuration issues have been the root cause for several recent attacks on IoT devices. For example, Mirai exploited the use of default credentials to infect IoT devices~\cite{antonakakis:2017:Mirai},  
while certain BLE-enabled baby monitors, vaping products and e-skateboards did not implement basic protection for their data~\cite{CVE:2019:CVE-2019-16518, CVE:2018:CVE-2018-10825, Zetter:2015:eskateboard}. This left the devices vulnerable to false data injection and the potential for physical harm to the user. 
The configuration of an IoT device can therefore reveal exploitable vulnerabilities and is a vector of interest in security analyses. 

There are several possible sources for this configuration information, including the devices themselves, the firmware they run, or any application or website they interface with. We analyse the merits and shortcomings of each of these potential information sources below.

\begin{figure*}[ht!]
\begin{subfigure}{\textwidth}
\begin{lstlisting}
 uint8_t passkey[] = "123456"; 
 ble_opt_t ble_opt; 
 ble_opt.gap_opt.passkey.p_passkey = &passkey[0]; 
 err_code = sd_ble_opt_set(BLE_GAP_OPT_PASSKEY, &ble_opt);
\end{lstlisting}
\vspace*{-3mm}
\caption{Source C code.}
\label{fig:fblock-code0}
\end{subfigure}

\vspace*{3mm}

\begin{subfigure}{0.525\textwidth}
\input{code/intro-example/example_1}
\vspace*{-3mm}
\caption{Disassembly of unstripped binary.}
\label{fig:fblock-code1}
\end{subfigure}
\hfill
\begin{subfigure}{0.45\textwidth}

\input{code/intro-example/example_2}
\vspace*{-3mm}
\caption{Disassembly of stripped binary.}
\label{fig:fblock-code2}
\end{subfigure}

\vspace*{-1mm}
\caption{Differences in the disassembly of unstripped and stripped binaries.}
\end{figure*}

\vspace{0.1cm}

\textbf{Devices:} Several security and privacy analyses have been conducted against IoT devices~\cite{sachidananda:2019:iotDos, loi:2017:IoT, das:2016:blePrivacy, Sivakumaran:2018:lowprofile}. Interfacing with physical devices can reveal behavioural characteristics, particularly those where user interaction is required. 
Combining hardware device tests with an analysis of communication interfaces can yield even more details. 
However, large-scale analyses can be difficult to automate as well as being prohibitively expensive due to the need for purchasing devices. 
Further, a variety of communication protocols may be used, particularly with IoT peripherals, which could require specialist hardware or software for each traffic analysis. 
    
\textbf{Mobile applications:} IoT devices, particularly those that are resource-constrained, commonly interface with mobile applications. Such mobile apps are often available as large repositories, are reasonably easy to analyse, and can provide indications of higher-layer processing. As such, they have been used in security analyses to identify vulnerabilities in the associated devices~\cite{chen:2018:iotfuzzer, chen:2016:firmadyne, wang:2019:iotMobileApps, Sivakumaran:2019:Crossapp, zuo:2019:uuid}. However, one app may interface with multiple devices, making it difficult to separate out relevant information for a single device. In addition, low-level protocol details may occur at the mobile operating system level, transparent to the app, or the app may act as a conduit between a device and a server without processing the data itself. In such cases, it may not be possible to get a complete picture of the IoT device's security via the mobile app.

\textbf{Web interface:} If an IoT device communicates with an external server, the exchanged messages may reveal some information regarding its configuration, particularly if it receives configuration commands from the server. However, performing tests against these servers may have legal implications. Also, large-scale automated tests may not always be feasible in the absence of physical devices because the server may require authenticated requests from the device~\cite{zhou:2019;iot}.  
    
\textbf{Firmware:} The firmware on an IoT device tends to apply to a single type of device (unlike mobile apps where one app may communicate with multiple devices) and generally reflects its configuration and functionality exactly. This  has lead firmware binaries to be the information source of choice for a number of security analyses~\cite{keliris:2018:icsref, costin:2014:embeddedfirmware, costin:2016:automated, shoshitaishvili:2015:firmalice, davidson:2013:fie}. However, firmware binaries are not always easy to obtain, as developers do not always make them publicly available. But more importantly, firmware analysis is, due to its own nature, far more complex than, for example, mobile app analysis.

\vspace{0.1cm}

The above analysis shows that, arguably, device firmware is the richest standalone source of information regarding a device's security configuration. Unfortunately, analysing firmware binaries is generally not straightforward. This is true even for devices such as IoT hubs and gateways (e.g., mobile phones and routers), which often run some version of the Linux operating system and therefore have familiar filesystem structures and commands that may be identifiable within their binaries. Analysis is much more complex for IoT \textit{peripherals}, which may run on top of custom operating systems, or sometimes have no operating system at all.
In this paper, we focus on peripheral firmware analysis.

IoT peripherals may implement one or more communication technologies, such as BLE~\cite{BluetoothSIG:2019:IntroBLE}, Zigbee~\cite{Zigbee:2019:IntroZigbee}, ANT~\cite{ANT:2019:IntroANT} or Thread~\cite{Thread:2019:IntroThread}. 
Many of these technologies have fully-fledged stacks, with protocols defined from the physical up to application layers. 
To reduce development time and ease application development, many IoT SoC vendors implement the technology stacks themselves and provide APIs through which application developers can incorporate some functionality provided by the stacks~\cite{Nordic:2020:SoftDevice, TI:2020:ZigbeeStack, TI:2020:BLEStack}.
In addition, developers may use library functions to perform other configurations. 
We present an example configuration function using sample C code in Figure~\ref{fig:fblock-code0}, where a fixed passkey is defined for the BLE pairing process using an API call \texttt{sd\_ble\_opt\_set}. We will use this example throughout the paper. 

Focusing on this example, it is known that fixed passkeys are a vulnerability, as they reduce the security of the pairing mechanism. A security analyst would therefore want to identify such uses of fixed passkeys from devices' firmware (since in most cases, the source code will not be publicly available). That is, they would want to know whether \texttt{sd\_ble\_opt\_set} (which we term a \textit{Call of Interest}, or COI) is called with a fixed passkey as its argument. To do this, we would need to pinpoint the location of the function call within the firmware binary, and then analyse the arguments that are passed to it. 
Figure~\ref{fig:fblock-code1} depicts the assembly instructions corresponding to this section of code, obtained by disassembling the firmware binary. 
From the instructions, we are able to identify that the relevant function call occurs at address \texttt{0x1ead0}, that the passkey bytes occur at address \texttt{0x21f14}, and that they are referenced by their absolute location at address \texttt{0x1ed9c}. 

The ability to correctly deduce the above pieces of information depends on a set of  \textbf{conditions}: 
\begin{enumerate}[label=C\arabic*,noitemsep]
    \item \label{item:condition1} Knowledge of function location and callers' addresses (i.e., knowing that the code for \texttt{sd\_ble\_opt\_set} is at address \texttt{0x1e748} and that it is called at address \texttt{0x1ead0}).
    \item \label{item:condition2} Knowledge of locations of inline data (i.e., knowing that the bytes at addresses \texttt{0x1ed9c} and \texttt{0x21f0c} should be interpreted as data rather than as code).
    \item \label{item:condition3} Firmware being loaded at the correct address (such that the absolute address \texttt{0x21f14} results in bytes being loaded from the correct location).
\end{enumerate}

This information is present within headers and symbol tables within the firmware. However, due to storage considerations, most IoT peripherals tend to ship firmware with this information removed, i.e., as \textit{stripped binaries}. 

Figure~\ref{fig:fblock-code2} depicts the disassembly of the binary file with ELF headers and debugging symbols stripped out. 
The disassembly of the stripped binary does not contain information about function names, thereby making it difficult to pinpoint locations of function calls (failing Condition~\ref{item:condition1}). 
Data segments have been incorrectly interpreted by the disassembler as code (failing Condition~\ref{item:condition2}), which leads to incorrect results when performing value tracing and precludes the use of emulation frameworks (e.g., QEMU~\cite{bellard:2005:qemu}, unicorn~\cite{Quynh:2020:Unicorn}). Further, the code has been loaded at the incorrect offset (failing Condition~\ref{item:condition3}), which means absolute addressing will fail.

Contributing to this problem is the fact that many resource-constrained IoT devices feature ARM processors~\cite{mishra:2015:wearable} with the Thumb and Thumb-2 instruction sets,\footnote{The Thumb instruction set has 2- and 4-byte wide instructions, unlike the fixed 4-byte wide instructions used in the ARM instruction set.} which offer greater code densities~\cite{goudge:1996:thumb}. In fact, the ARM \mbox{Cortex-M} processors, which are very popular in embedded systems, support \textit{only} the Thumb and Thumb-2 instruction sets. These instruction sets are not yet fully supported by many disassemblers.

\IncMargin{1.0em}
\begin{algorithm}[t!]
\caption{Application code base identification}
\footnotesize
\label{algo:app-code-base}
\Indm 
\KwResult{Application code base}
\Indp
\BlankLine
$vtAddresses = []$ \Comment*{\footnotesize Vector Table (VT) addresses.} 
\For{$vtIndex \in [1,2,3,4,5,6,14,15]$}{
    $vtEntry = readBytesFromBinary(vtIndex,vtIndex+4)$\;
    $vtAddresses.insert(vtEntry-1)$\;
}

\For{$branchIns \in disassembledInstructions$}{
    \If{$target(branchIns) == address(branchIns)$}{
        \For{$vtAddress \in vtAddresses$}{
            \Comment{\footnotesize Consider addresses whose last 3 hex digits match those of a VT entry.} 
            \If{$vtAddress[-3:] == address(branchIns)[-3:]$}{
                $appCodeBase = (vtAddress - address(branchIns))$\;
            }
        }
    }
}
 
\end{algorithm}
\DecMargin{1.0em}

By testing different reverse-engineering tools against stripped \mbox{Cortex-M} IoT binaries, we found that \texttt{IDA} (free)~\cite{hex:2010:ida} does not currently support ARM, while \texttt{Debin}~\cite{he:2018:Debin} and \texttt{BAP}~\cite{brumley:2011:bap} do not fully support the Thumb instruction set. 
In addition, \texttt{radare2}~\cite{radare:2020:radare2} exhibited stability issues, while \texttt{angr}~\cite{wang:2017:angr} was unable to produce a valid Control Flow Graph (CFG) - a step prior to analysis - for stripped \mbox{Cortex-M} binaries. 
Our observation regarding the robustness of \texttt{angr} and \texttt{radare2} for Thumb mode analysis is supported by~\cite{Jiang:2020:ArmDisassemblyTools}, which also noted that \texttt{Ghidra}~\cite{NSA:2020:Ghidra} too has better support for the ARM instruction set than for Thumb. 
Further, because IoT peripheral binaries typically do not include the technology stack or ROM data, \textit{dynamic} analysis approaches are unsuitable. This reveals a gap in the automated IoT security analysis landscape and prompted the development of \texttt{argXtract}.

\section{\texttt{argXtract}}
\label{section:system-description}
\texttt{argXtract} begins with assembly code for a stripped binary file, obtained by disassembling the binary using Capstone in ARM Thumb mode. It processes the disassembled instructions to overcome several unhandled instances of data misinterpreted as code, prior to pinpointing configuration-related \textit{Calls of Interest (COIs)} and extracting the arguments passed to them.
It supports the entire ARMv6-M instruction set and the most prevalent instructions within ARMv7-M.

In this section, we describe the following aspects of \texttt{argXtract}: \S\ref{subsection:app-code-base} - Application code base identification, for correct absolute addressing; \S\ref{subsection:data-annotation} - Inline data identification, such that data is not incorrectly interpreted as code; \S\ref{subsection:id-function-blocks} -  Function block identification, to enable call execution path generation and to enable function pattern matching; \S\ref{subsection:coi-identification} - COI (ARM supervisor call or function call) identification, to produce a list of trace termination points; \S\ref{subsection:arg-matching} - Tracing and argument processing, to determine the input arguments passed to a COI.

\IncMargin{1.0em}
\begin{algorithm}[t!]
\caption{Inline data identification (\texttt{ldr})}
\footnotesize
\label{algo:inline-data-ldr}
\For{$instruction \in disassembledInstructions$}{
    \If{$opcode(instruction) \in ldrInstructions$}{
        $ldrTarget = target(instruction)$\;
        \If{$isPcRelativeAddress(ldrTarget)$}{
            \uIf{$opcode(instruction) == ldr$}{
                $numBytes = 4$\;
            }
            \uElseIf{$(opcode(instruction) == ldrh) | (opcode(instruction) == ldrsh)$}{
                $numBytes = 2$\;
            }
            \Else{
                $numBytes = 1$\;
            }
            $markAddressAsData(ldrTarget,numBytes)$\;
            \If{$(numBytesAtTarget == 4) \& (numBytes < 4)$}{
                $reinterpretOverflowAsInstruction(ldrTarget+2)$\;
            }
        }
        
    }
}
 
\end{algorithm}
\DecMargin{1.0em}

\subsection{Application Code Base Identification} \label{subsection:app-code-base}

As described in \S\ref{section:background}, using an incorrect offset for instruction addresses will lead to the failure of absolute addressing. 
\texttt{argXtract} combines \textit{known} address information with \textit{obtained} addresses to compute the application code base using Algorithm~\ref{algo:app-code-base}. 
The addresses of core interrupt handlers are known, as they are present at specific offsets (line 2) within the Application Vector Table (VT), which is located at \texttt{0x00000000} within the stripped binary~\cite{ARM:2020:VectorTable}. 
The addresses will have a Least Significant Bit (LSB) of 1, indicating a switch to Thumb mode. The actual address is obtained by subtracting 1 (line 4). 
Corresponding interrupt handler \textit{code} within the stripped binary is identified by exploiting the fact that at least one interrupt handler is usually the \textit{default handler}, i.e., an endless loop or self-targeting branch.
\texttt{argXtract} extracts all self-targeting branches (lines 6-7) and compares their addresses against the VT addresses to compute the app code base (lines 8-10).
Once the application code base has been identified, the binary disassembly is reloaded at the correct offset. It now satisfies Condition~\ref{item:condition3} as presented in \S\ref{section:background}. 

\subsection{Inline Data Identification}
\label{subsection:data-annotation}

\begin{figure}[t!]
\begin{lstlisting}
 2894a:	2e08        cmp	r6, #8
 2894c:	d219        bcs.n  28982
 2894e:	e8df f006   tbb [pc, r6]
 28952:	1b181804    ; data
 28956:	172f2f22    ; data
 2895a:	8901        ldrh r1, [r0, #8]
\end{lstlisting}
\caption{Sample table branch structure.}
  \label{fig:tbb}
\end{figure}
\IncMargin{1.0em}
\begin{algorithm}[t!]
\caption{Inline data identification (\texttt{tbb}, \texttt{tbh})}
\footnotesize
\label{algo:inline-data-tbb}
\For{$instruction \in disassembledInstructions$}{
    \If{$opcode(instruction) \in tableBranchInstructions$}{
        $mulFactor = (opcode(instruction) == tbh) ? 2 : 1$\;
        $cmpValue = getTableSkipComparisonValue(address(instruction))$\;
        $maxBranchAddress = pcAddress + (cmpValue * mulFactor) + 2$\;
        \If{$(opcode(instruction) == tbb) and (byte(maxBranchAddress) == 00)$}{
            $maxBranchAddress \mathrel{+}= 1$\;
        }
        $tableBranchAddress = pcAddress$\;
        \While{$tableBranchAddress < maxBranchAddress$}{
            $markAddressAsData(tableBranchAddress)$\;
            $tableBranchAddress \mathrel{+}= 2$\;
        }
    }
    
}
 
\end{algorithm}
\DecMargin{1.0em}

Stripped \mbox{Cortex-M} binaries do not contain section information.
Their disassembly therefore produces a block of instructions with a \verb+.text+ (i.e., code) segment and often a \verb+.data+ segment, with no demarcation between the two and the \verb+.data+ segment misinterpreted as code.
The \verb+.text+ segment also has inline data, often misinterpreted as code and resulting in value tracing errors. 

\texttt{argXtract} uses information from the Reset Handler, whose address is read from the Application Vector Table, to identify the location and correct starting address of the \verb+.data+ segment. It then identifies inline data using three primary sources: (i) PC-relative memory-loads (e.g., \texttt{ldr}, \texttt{ldrh}), (ii) table branches (\texttt{tbb}, \texttt{tbh}) and compact switch table helpers such as \texttt{\_\_ARM\_common\_switch8} and \texttt{\_\_gnu\_thumb1} variants, and (iii) direct write-to-PC operations. These operations aid in satisfying Condition~\ref{item:condition2} (described in \S\ref{section:background}). 
We describe the inline data identification mechanism for each of these sources in further detail below.

\paragraph{Identification of \texttt{.data}}
The Reset Handler often contains the final address of the \texttt{.text} segment as well as the start and end addresses for the \texttt{.data} segment. This is present in the form of consecutive memory-loads. \texttt{argXtract} extracts source addresses from consecutive memory-load instructions within the Reset Handler code and analyses them to determine whether they match the required structure. If they do, then the addresses starting after the final address of the \texttt{.text} segment and ending at the end of the file are marked as data.

\paragraph{PC-relative memory-loads}
A memory-load (i.e., \texttt{ldr} and variants) that loads data from an address within the firmware file will specify the source address relative to either the Program Counter (PC) or a register. Register-relative loads may require significant tracing in some cases. However, PC-relative loads are straightforward to analyse.
\texttt{argXtract} performs a linear scan for PC-relative memory-loads, calculates the address from which data is loaded and marks it as data, re-processing residual bytes as instructions where required. This is described in Algorithm~\ref{algo:inline-data-ldr}.

\paragraph{Table branches}
Table branch instructions (\texttt{tbb}, \texttt{tbh}) were introduced in the ARMv7-M architecture to handle complex branching conditions. 
Figure~\ref{fig:tbb} depicts a sample table branch instruction (at address \texttt{0x2894e}). 
The instruction is immediately followed by a branch table (\texttt{0x28952} and \texttt{0x28956}). 
This table should be interpreted as data, but is misinterpreted by disassemblers as code in the absence of debugging symbols. 
\texttt{argXtract} follows the procedure described in Algorithm~\ref{algo:inline-data-tbb}, exploiting the comparison made with the indexing register (\texttt{0x2894a}) to identify the range of addresses that make up the branch table and mark it as data.

\begin{figure}
\begin{lstlisting}
 1a83c:	2c17    cmp    r4, #23
 1a83e:	d8fc    bhi.n  1a83a
 1a840:	4ac7    ldr    r2, [pc, #796]
 1a842:	00a3    lsls   r3, r4, #2
 1a844:	58d3    ldr    r3, [r2, r3]
 1a846:	469f    mov    pc, r3
\end{lstlisting}
\caption{Write-to-PC operation.}
  \label{fig:write-to-pc}
\end{figure}

\paragraph{Compact switch helpers}
Prior to the introduction of table branch instructions, helper functions were utilised to handle switch-case constructs. The GCC compiler produces \texttt{\_\_gnu\_thumb1} variants, while Keil produces \texttt{\_\_ARM\_common\_switch8}. 
These helper functions have identifiable function prologues, and calls to the functions are followed by an index table, similar to table branch instructions. 
\texttt{argXtract} determines the locations of helper functions and applies function-specific processing to determine the size of the index table. It also determines the addresses of resultant branches, to be used for function block identification (\S\ref{subsection:id-function-blocks}).

\IncMargin{1.0em}
\begin{algorithm}[t!]
\caption{Function estimation}
\footnotesize
\label{algo:func-estimation}
\Indm 
\KwResult{Start addresses for estimated function blocks}
\Indp
\BlankLine
$functionAddresses = [vtAddresses, branchTargets].sort()$\;
\For{$functionAddress \in functionAddresses$}{
    $start = functionAddress$\;
    $end = getFinalAddressInBlock(functionAddress)$\;
    $exitPoints = getExitPoints(start,end).sort()$ 
    
    \For{$i = start \to end$}{
        \uIf{$opcode(i) \in conditionalBranchInstr$}{
            
            $branchTarget_i = target(i)$ \Comment*[r]{\footnotesize Target of branch.} 
            \For{$exit \in exitPoints$}{
                \If{$doesBranchBypassExit(branchTarget_i, exit)$}{
                    $exitPoints.remove(exit)$\;
                }
            }
        }
         \ElseIf {$i \in [switchCalls, PCwrites]$}{
           $maxBranchAddress = max(switchTable)$\;
           \If{$doesOverflowBlock (maxBranchAddress,  end)$} {
                $combineAndReestimate(maxBranchAddress, start)$\;
            }
           \For{$exit \in exitPoints$}{
                \If{$doesBypassExit(maxBranchAddress, exit)$}{
                    $exitPoints.remove(exit)$\;
                }
            }
        }
    }
    
    $nextFunctionStart = getFirstInstructionPostExit(exitPoints)$\;
    $functionAddresses.insert(nextFunctionStart))$\;
} 
\end{algorithm}
\DecMargin{1.0em}

\paragraph{Write-to-PC operations}
Direct write-to-PC operations are sometimes used to accomplish code branches. 
Figure~\ref{fig:write-to-pc} depicts an example.
This operation loads a branch address from an address within the firmware and writes the branch address to the PC. The address from which the branch address is loaded (i.e., the \texttt{ldr} source at \texttt{0x1a844}, obtained in this example by adding the contents of \texttt{r2} and \texttt{r3}) must be interpreted as data, but is misinterpreted as code when disassembling stripped binaries.

When a write-to-PC is encountered (at 0x1a846 in Figure~\ref{fig:write-to-pc}), \texttt{argXtract} examines the preceding instructions until an integer comparison is identified (0x1a83c). It then uses subsequent conditional branches (0x1a83e) to determine the actual range of values for the comparison register ([0,23] in this example). The instructions following the branch and until the non-PC-relative memory-load (0x1a844) are executed for all possible values of the comparison register. This produces a range of addresses from which the branch addresses are loaded. The range of addresses is marked as data. A second execution until the PC-write instruction produces the set of branch addresses, which are stored to be used in a subsequent step by the function identification component (\S\ref{subsection:id-function-blocks}).

\begin{figure}[!t]
\begin{lstlisting}
  18228:  bl  182b0 <functionB>

000182b0 <functionB>:
  182b0:  push  {r4, lr}
  182b2:  cmp  r2, #32
  182b4:  blt.n  182c0  (*\small;skips pop at 182be*)
  182b6:  mov  r0, r1
  182b8:  subs  r2, #32
  182ba:  lsrs  r0, r2
  182bc:  movs  r1, #0
  182be:  pop  {r4, pc}
  182c0:  mov  r3, r1
  182c2:  lsrs  r3, r2
  182c4:  lsrs  r0, r2
  182c6:  movs  r4, #32
  182c8:  subs  r2, r4, r2
  182ca:  lsls  r1, r2
  182cc:  orrs  r0, r1
  182ce:  mov  r1, r3
  182d0:  pop  {r4, pc}
  182d2:  nop

000182d4 <functionC>:   (*\small;called indirectly*)
  182d4:  ldrb  r2, [r0, #0]
  182d6:  adds  r0, r0, #1
\end{lstlisting}
\caption{Reference ARM assembly}
  \label{fig:asm-function}
\end{figure}

\subsection{Function Block Identification} \label{subsection:id-function-blocks}

The challenges involved in identifying function blocks within stripped binaries have been widely studied. Indirect function calls, absence of specific function prologues, indeterminate location of start instructions, absence of a clear exit point or presence of multiple exit points are some of the problems that have been outlined in the literature~\cite{yin:2018:functionstrippedbinary} and which we have observed with regard to function block identification.

\begin{figure*}[!t]
\begin{subfigure}{0.32\textwidth}
\begin{lstlisting}[basicstyle=\footnotesize]
0x18228:  bl  0x182b0
...
0x182b0:  push  {r4, lr}
0x182b2:  cmp  r2, #0x20
0x182b4:  blt  #0x182c0
0x182b6:  mov  r0, r1
0x182b8:  subs  r2, #0x20
0x182ba:  lsrs  r0, r2
0x182bc:  movs  r1, #0
0x182be:  pop  {r4, pc}
0x182c0:  mov  r3, r1
0x182c2:  lsrs  r3, r2
0x182c4:  lsrs  r0, r2
0x182c6:  movs  r4, #0x20
0x182c8:  subs  r2, r4, r2
0x182ca:  lsls  r1, r2
0x182cc:  orrs  r0, r1
0x182ce:  mov  r1, r3
0x182d0:  pop  {r4, pc}
0x182d2:  nop
0x182d4:  ldrb  r2, [r0]
0x182d6:  adds  r0, r0, #1
\end{lstlisting}
\vspace{-2mm}
\caption{Capstone disassembly (starting point).}
\label{fig:fblock-capstone0}
\end{subfigure}
\hfill 
\begin{subfigure}{0.32\textwidth}
\begin{lstlisting}[basicstyle=\footnotesize]
0x18228:  (*\bfseries bl  0x182b0*)
...
(*\bfseries {\color{cb1}0x182b0: \enspace\enspace\enspace\enspace  push  {r4, lr} ;start} *)
0x182b2:  cmp  r2, #0x20
0x182b4:  blt  #0x182c0
0x182b6:  mov  r0, r1
0x182b8:  subs  r2, #0x20
0x182ba:  lsrs  r0, r2
0x182bc:  movs  r1, #0
0x182be:  pop  {r4, pc}
0x182c0:  mov  r3, r1
0x182c2:  lsrs  r3, r2
0x182c4:  lsrs  r0, r2
0x182c6:  movs  r4, #0x20
0x182c8:  subs  r2, r4, r2
0x182ca:  lsls  r1, r2
0x182cc:  orrs  r0, r1
0x182ce:  mov  r1, r3
0x182d0:  pop  {r4, pc}
0x182d2:  nop
0x182d4:  ldrb  r2, [r0]
0x182d6:  adds  r0, r0, #1
\end{lstlisting}
\vspace{-2mm}
\caption{Identify function blocks using \texttt{b}, \texttt{bl}.}
\label{fig:fblock-capstone1}
\end{subfigure}
\hfill 
\begin{subfigure}{0.32\textwidth}
\begin{lstlisting}[basicstyle=\footnotesize]
0x18228:  bl  0x182b0
...
(*\bfseries {\color{cb1}0x182b0: \enspace\enspace\enspace\enspace  push  {r4, lr} ;start} *)
0x182b2:  cmp  r2, #0x20
0x182b4:  blt  #0x182c0
0x182b6:  mov  r0, r1
0x182b8:  subs  r2, #0x20
0x182ba:  lsrs  r0, r2
0x182bc:  movs  r1, #0
(*\bfseries {\color{contrastcol4}0x182be: \enspace\enspace\enspace\enspace pop  \{r4, pc\} ;end?} *)
0x182c0:  mov  r3, r1
0x182c2:  lsrs  r3, r2
0x182c4:  lsrs  r0, r2
0x182c6:  movs  r4, #0x20
0x182c8:  subs  r2, r4, r2
0x182ca:  lsls  r1, r2
0x182cc:  orrs  r0, r1
0x182ce:  mov  r1, r3
(*\bfseries {\color{contrastcol4}0x182d0: \enspace\enspace\enspace\enspace pop  \{r4, pc\} ;end?} *)
0x182d2:  nop
0x182d4:  ldrb  r2, [r0]
0x182d6:  adds  r0, r0, #1
\end{lstlisting}
\vspace{-2mm}
\caption{Mark potential exit points such as \texttt{pop} \texttt{pc}}
\label{fig:fblock-capstone2}
\end{subfigure}

\vspace{2mm}

\begin{subfigure}{0.32\textwidth}
\begin{lstlisting}[basicstyle=\footnotesize]
0x18228:  bl  0x182b0
...
(*\bfseries {\color{cb1}0x182b0: \enspace\enspace\enspace\enspace  push  {r4, lr} ;start} *)
0x182b2:  cmp  r2, #0x20
(*\bfseries {\color{contrastcol3}0x182b4: \enspace\enspace\enspace\enspace blt  \#0x182c0} *)
0x182b6:  mov  r0, r1
0x182b8:  subs  r2, #0x20
0x182ba:  lsrs  r0, r2
0x182bc:  movs  r1, #0
(*\bfseries {\color{contrastcol4}0x182be: \enspace\enspace\enspace\enspace pop  \{r4, pc\} ;end?} *)
0x182c0:  mov  r3, r1
0x182c2:  lsrs  r3, r2
0x182c4:  lsrs  r0, r2
0x182c6:  movs  r4, #0x20
0x182c8:  subs  r2, r4, r2
0x182ca:  lsls  r1, r2
0x182cc:  orrs  r0, r1
0x182ce:  mov  r1, r3
(*\bfseries {\color{contrastcol4}0x182d0: \enspace\enspace\enspace\enspace pop  \{r4, pc\} ;end?} *)
0x182d2:  nop
0x182d4:  ldrb  r2, [r0]
0x182d6:  adds  r0, r0, #1
\end{lstlisting}
\vspace{-2mm}
\caption{Identify instructions that skip exit points.}
\label{fig:fblock-capstone3}
\end{subfigure}
\hfill %
\begin{subfigure}{0.32\textwidth}
\begin{lstlisting}[basicstyle=\footnotesize]
0x18228:  bl  0x182b0
...
(*\bfseries {\color{cb1}0x182b0: \enspace\enspace\enspace\enspace  push  {r4, lr} ;start} *)
0x182b2:  cmp  r2, #0x20
0x182b4:  blt  #0x182c0
0x182b6:  mov  r0, r1
0x182b8:  subs  r2, #0x20
0x182ba:  lsrs  r0, r2
0x182bc:  movs  r1, #0
0x182be:  pop  {r4, pc} 
0x182c0:  mov  r3, r1
0x182c2:  lsrs  r3, r2
0x182c4:  lsrs  r0, r2
0x182c6:  movs  r4, #0x20
0x182c8:  subs  r2, r4, r2
0x182ca:  lsls  r1, r2
0x182cc:  orrs  r0, r1
0x182ce:  mov  r1, r3
(*\bfseries {\color{contrastcol4}0x182d0: \enspace\enspace\enspace\enspace pop  \{r4, pc\} ;end?} *)
0x182d2:  nop
0x182d4:  ldrb  r2, [r0]
0x182d6:  adds  r0, r0, #1
\end{lstlisting}
\vspace{-2mm}
\caption{Remove skipped exit instructions.}
\label{fig:fblock-capstone4}
\end{subfigure}
\hfill %
\begin{subfigure}{0.32\textwidth}
\begin{lstlisting}[basicstyle=\footnotesize]
0x18228:  bl  0x182b0
...
(*\bfseries {\color{cb1}0x182b0: \enspace\enspace\enspace\enspace  push  {r4, lr} ;start} *)
0x182b2:  cmp  r2, #0x20
0x182b4:  blt  #0x182c0
0x182b6:  mov  r0, r1
0x182b8:  subs  r2, #0x20
0x182ba:  lsrs  r0, r2
0x182bc:  movs  r1, #0
0x182be:  pop  {r4, pc}
0x182c0:  mov  r3, r1
0x182c2:  lsrs  r3, r2
0x182c4:  lsrs  r0, r2
0x182c6:  movs  r4, #0x20
0x182c8:  subs  r2, r4, r2
0x182ca:  lsls  r1, r2
0x182cc:  orrs  r0, r1
0x182ce:  mov  r1, r3
0x182d0:  pop  {r4, pc}
0x182d2:  nop
(*\bfseries {\color{cb1}0x182d4: \enspace\enspace\enspace\enspace ldrb  r2, [r0] ;start }*)
0x182d6:  adds  r0, r0, #1
\end{lstlisting}
\vspace{-2mm}
\caption{Identify suitable start for next function.}
\label{fig:fblock-capstone5}
\end{subfigure}

\caption{Process used by \texttt{argXtract} for identifying function blocks.}
\end{figure*}

\texttt{argXtract} performs function block identification to be able to enable function pattern matching and call execution path determination. 
To begin function block estimation, \texttt{argXtract} first produces an initial set of high-certainty candidates for function start addresses by extracting the addresses of interrupt handlers from the Application Vector Table. 
That is, each interrupt handler is considered as a separate function and the addresses of the interrupt handler functions are obtained from the VT that occurs at the beginning of the binary file.
Targets of branch-and-link (\texttt{bl}) instructions are added to this set; targets of branch (\texttt{b}) instructions are also added, but subject to satisfying requirements regarding function prologues (specifically, the instruction at the branch target address must be a \texttt{push}, \texttt{sub} \texttt{sp}, or \texttt{stmdb} \texttt{sp[!]}, \texttt{\{..,lr\}}).

A function estimation algorithm (Algorithm~\ref{algo:func-estimation}) is then executed against each block of instructions beginning from one start address and ending prior to the next start. The algorithm operates on the basic principle that, while a function may have multiple exit instructions due to conditional executions, it must have mechanisms for bypassing all but one of the exit points. This could be via conditional branch instructions or a switch/branch table (as identified in \S\ref{subsection:data-annotation}). \texttt{argXtract} determines all potential exit points (\texttt{pop}, \texttt{bx} \texttt{lr}, unconditional branches to lower addresses or outside the current block, and data) within the block of instructions that is being analysed and marks the exit point that \textit{cannot be bypassed} as the ultimate function exit. The next valid instruction is determined to be the beginning of the next function. This procedure is performed iteratively to obtain the final list of function blocks.

We further illustrate this algorithm using the code example in Figure~\ref{fig:asm-function} as reference. This reference code contains two functions, denoted as \texttt{functionB} and \texttt{functionC}. Of these, \texttt{functionB} gets called via a \texttt{bl} instruction within code, while \texttt{functionC} does not. \texttt{functionC} gets called indirectly via a \texttt{blx} call, which cannot be identified without some level of register tracing. 

Figure~\ref{fig:fblock-capstone0} depicts the equivalent assembly code obtained using Capstone against the stripped version of the binary. We apply Algorithm~\ref{algo:func-estimation} to this as follows:
\begin{itemize}[noitemsep]
    \item Identify potential function starts, from targets of \texttt{bl} and \texttt{b} instructions. Figure~\ref{fig:fblock-capstone1} shows that there is one such start, at address \texttt{0x182b0}. This corresponds to \texttt{functionB}.
    \item Mark out possible exit points, such as \texttt{pop} \texttt{pc}, \texttt{bx} \texttt{lr}, or data. As shown in Figure~\ref{fig:fblock-capstone2}, there are two such potential exit points, at addresses \texttt{0x182be} and \texttt{0x182d0}.
    \item Look for branches that skip the exit points. Figure~\ref{fig:fblock-capstone3} shows that the branch condition at \texttt{0x182b4} skips the exit point at \texttt{0x182be}. This exit point is therefore considered to be part of the existing function block (i.e., the one beginning at \texttt{0x182b0}).
    \item Remove exit points that are skipped. We are left with one other potential exit point at \texttt{0x182d0}. It is considered the final exit point of the function block (see Figure~\ref{fig:fblock-capstone4}).
    \item Identify the next valid instruction as the start of the next function block. Initially, we consider the instruction at \texttt{0x182d2} to be the start of the next function block. However, functions tend not to begin with \texttt{nop} instructions. Therefore, we skip past this and mark \texttt{0x182d4} as the start of the next function block, as shown in Figure~\ref{fig:fblock-capstone5}. 
\end{itemize}

 Comparing the obtained function address with Figure~\ref{fig:asm-function}, we can see that the algorithm has correctly identified the function block starting address for \texttt{functionC}.

This process is then repeated from the start of the new function block (i.e., from \texttt{0x182d4}), until the end of the block is reached.
At the end of this step, we have a list of function start addresses (i.e., the addresses of functions) for the binary.


\paragraph{Function block annotation}
\texttt{argXtract} maintains a function block object, with a key for each identified function block. 
The object contains information on cross-references to and from a function block, as well as the \textit{call-depth}. The call-depth indicates the maximum number of functions that get called iteratively by a function. 
This tells \texttt{argXtract} which functions will take a long time to execute and are therefore candidates for exclusion when execution is time-limited. 
If a function contains perpetual loops, as is the case with some error handlers, then it is deny-listed, to prevent inadvertently calling such functions and causing the tool to execute forever.

\subsection{COI Identification}
\label{subsection:coi-identification}
\texttt{argXtract} identifies calls to API methods, which could be standard function calls or which could be translated to ARM supervisor calls. Supervisor calls and function calls are analysed using distinct techniques, as described in \S\ref{subsubsection:svc-id} and \S\ref{subsubsection:pattern-matching}, respectively. This then satisfies Condition~\ref{item:condition1} as described in \S\ref{section:background}. After this step, all pre-conditions for analysing a stripped ARM binary will be satisfied.

\subsubsection{Supervisor Call Identification}
\label{subsubsection:svc-id}
An ARM supervisor call is simply an instruction with an opcode of \texttt{svc}. 
A supervisor call will have an associated \texttt{svc} number.
If an IoT chipset vendor issues a technology stack that accepts configuration commands via supervisor calls, then the associated \texttt{svc} numbers will normally be available from the vendor SDK.

In \texttt{svc} analysis mode, an input object containing the \texttt{svc} numbers of interest is provided to \texttt{argXtract}. The tool performs a linear scan over the disassembly for \texttt{svc} instructions and tests the \texttt{svc} numbers against those that are provided to it. It stores the addresses of relevant \texttt{svc} instructions, to be used in the tracing step (\S\ref{subsection:arg-matching}). 

\subsubsection{Function Pattern Matching}
\label{subsubsection:pattern-matching}
Identifying function calls of interest is far more complex than identifying supervisor calls, as functions cannot be immediately identified within assembly. \texttt{argXtract} exploits the fact that configuration API functions (such as those provided by vendors for performing configurations to IoT stacks) accept inputs in a specific order, which are passed within registers in a specific sequence (\texttt{r0}, \texttt{r1},...) for \mbox{Cortex-M}. Further, most functions generate artefacts that are detectable within memory and/or registers, i.e., as output or intermediate values. 

For each function of interest, we define a ``function pattern file'', which is a collection of test sets containing register and memory inputs, and the corresponding outputs (which could be actual output values, stored in registers or memory, or intermediate values at detectable locations). Many functions store identifiable values at binary-specific locations that \textit{cannot} be predetermined. These are handled using wildcard addresses, where expected values are specified at some predetermined offset from the wildcard address.

The function pattern files are passed to each of the functions that have been identified for the binary under test (using the process described in \S\ref{subsection:id-function-blocks}). The function instructions are executed with the input register and memory values specified in the pattern file. Output register and memory contents are compared against the expected values. 
If a single function matches the given pattern, then this is taken to be the function of interest. In the case of nested function calls, the function with the lowest call depth that satisfies the given pattern is taken to be the function of interest. The starting address of the function and the addresses of calling instructions are stored, to be used during tracing.

We verified the pattern matching module against the \texttt{CryptoKeyPlaintext\_initKey} function from the Texas Instruments SimpleLink Platform, the \texttt{mbedtls\_ssl\_conf\_ciphersuites} from the mbedtls library, and the \texttt{ot::KeyManager::SetKeyRotation} OpenThread function. When testing for these functions, we generated stripped binaries using different vendor SDKs (where relevant), as well as different projects and compilers (Keil, IAR, Clang), to account for vendor- and compiler-introduced variations. \texttt{argXtract} was able to identify the correct function location from within the stripped binary in each case, which we verified using the unstripped versions of the binaries.

\paragraph{Reducing execution time with function matching}
In addition to matching configuration API functions, we also use function matching to map the identified function blocks to well-known, long-running functions. This reduces execution time when tracing as these mapped functions can be modelled. In our work, we have modelled the \texttt{memset} and \texttt{\_\_udivsi3} (integer division) functions.

\subsection{Register Tracing \& Argument Processing}\label{subsection:arg-matching}
Once COIs have been identified (as described in \S\ref{subsection:coi-identification}), \texttt{argXtract} performs backward inter-procedural tracing to determine all call execution paths leading to the COI(s). It then forward-traces along the paths while updating the values of registers, memory and conditional flags with each instruction execution. 
Note that \texttt{argXtract} does not, in the interest of processing time and resources, follow every possible branch that it encounters during the trace.\footnote{A fairly simple firmware binary can have several thousand branch instructions, which rapidly compounds the required memory and analysis time.} Instead, it follows a set of rules, to determine whether or not to proceed with a branch:
\begin{itemize}[noitemsep]
    \item All branches that are within the identified call execution path are followed.
    \item All unconditional internal loops (i.e., branches to other addresses within the same function block) are followed.
    \item Conditional internal loops are followed if the conditions are satisfied. If conditional checks cannot be made (e.g., due to the flags not being set as expected), then the outcome is \textit{indeterminate} and both possible paths are taken.
    \item Branches to deny-listed functions are not taken.
    \item Branches to function blocks that are \textit{not} present within the call execution path are only followed if their call-depth is less than a configurable $max\_call\_depth$ value (default=1). 
\end{itemize}

If a branch check should fail, then \texttt{argXtract} skips it and proceeds to the next instruction. If a \texttt{bl} is skipped, then going by the ARM convention of using register \texttt{r0} for holding the output of subroutines~\cite{ARM:2020:RegisterSubroutine}, and the convention of returning 0 on execution success, we artificially assign a value of 0 to \texttt{r0} in order to bypass a possible subsequent branch to fault handlers due to non-zero status returns.

Once the COI is reached via the trace, its arguments are extracted for processing. The arguments to a COI are contained within registers \texttt{r0}-\texttt{r3} (or are on the call stack)~\cite{wang:2017:embedded, ARM:2020:CallingSVC, ARM:2020:SVC}. 
Some registers may hold pointers to data in memory. 
Therefore, when a COI is reached, the contents of both the register object and the memory map (as shown in Figure~\ref{fig:arg-code-out-sample}) are returned to an argument analysis component for processing. 

\begin{figure}[!t]
\begin{lstlisting}[xleftmargin=0.08\columnwidth,xrightmargin=0.08\columnwidth]
 {
   "args": {
      "0": {...},
      "1": {
        "in_out": "in",
        "ptr_val": "pointer",
        "data": {
          "p_opt": {
            "ptr_val": "pointer",
            "length_bits": 48,
            "type": "hex"
          }
        }
      }
   }
 }
\end{lstlisting}
\vspace*{-3mm}
\caption{Argument definition object.}
  \label{fig:arg-def-sample}

\vspace*{3mm}

\begin{lstlisting}[xleftmargin=0.08\columnwidth,xrightmargin=0.08\columnwidth]
{"sd_ble_opt_set":{
  "memory":{..., 20007f60:"31", 20007f61:"32", 20007f62:"33", 20007f63:"34", 20007f64:"35", 20007f65:"36", 20007f66:"00", 20007f68:"60", 20007f69:"7f", 20007f6a:"00", 20007f6b:"20", ...}, 
  "registers":{..., "r0":"00000022", "r1":"20007f68", ...}}}
\end{lstlisting}
\vspace*{-3mm}
\caption{Register/memory contents.}
  \label{fig:arg-code-out-sample}

\vspace*{3mm}

\begin{lstlisting}[xleftmargin=0.08\columnwidth,xrightmargin=0.08\columnwidth]
 "output": {
    "sd_ble_opt_set": [
        {
            "opt_id": 34,
            "p_opt": "313233343536"
        }
    ]
 }
\end{lstlisting}
\vspace*{-3mm}
\caption{Processed output file.}
  \label{fig:arg-out-sample}

\end{figure}

The type and format of data that are used as arguments to COIs are obtained from header files within vendor SDKs and provided to \texttt{argXtract} in the form of \textit{Argument Definition Objects}. 
These are JSON files that describe the expected structure of data bits for each input argument using predefined keywords.\footnote{We adopt this template-based approach for greater flexibility, such that supporting additional COIs only requires including new Argument Definition Objects, rather than needing to add extra COI-specific code.} 
For example, Figure~\ref{fig:arg-def-sample} depicts the Argument Definition Object for the \texttt{sd\_ble\_opt\_set} COI discussed in \S\ref{section:background}. 
Taking the second argument as an example, we note that it is defined as a pointer to a pointer to a 6-byte (48-bit) array. 
This argument is contained within register \texttt{r1}, which according to the trace output in Figure~\ref{fig:arg-code-out-sample} contains a value of \texttt{20007f68}. 
According to the Argument Definition Object, this is a pointer, and the memory object in Figure~\ref{fig:arg-code-out-sample} shows that this memory address points to the address \texttt{20007f60}, which (also being a pointer) points to the hex value \texttt{0x313233343536}. 
This corresponds to the ASCII string ``123456'', which is the value specified as the fixed passkey in our example in Figure~\ref{fig:fblock-code0}. 
This results in the output file depicted in Figure~\ref{fig:arg-out-sample}.

Arguments that hold \textit{outputs} (by being written to a memory location) will be specified in the Argument Definition Object to be fed back into the memory map and used in subsequent traces. This is used to tie together two separate traces.
We use this functionality to link a BLE characteristic with its associated service when analysing Nordic binaries (\S\ref{section:nordic}).


\section{Evaluation}
\label{section:evaluation}
In this section, we evaluate \texttt{argXtract} in terms of the accuracy of function block identification and the correctness of the extracted configuration data.

\subsection{Test Set and Ground Truth}
There is no publicly available dataset of ARM \mbox{Cortex-M} binaries with known and annotated configurations. 
For this reason, we generated our own dataset, comprising 28 stripped binaries, for testing and verification purposes. The binaries target chipsets from NXP, STMicroelectronics, Nordic Semiconductor and Texas Instruments, for multiple IoT technologies including Zigbee, ANT, BLE, Thread and 802.15.4. The binaries were compiled using GCC, IAR, Keil and Clang, depending on the options made available by the chipset vendor. We provide a detailed description of the test set binaries in our repository (see \S\ref{section:resources}).

For ground truth, we obtained the actual configuration for each binary by disassembling its \textit{unstripped} version using the GNU ARM embedded toolchain.

\begin{table}[!t]
\centering
\footnotesize
\caption{True Positive Rates (TPR) and Effective False Positive Rates (EFPR) for function block identification against test binaries (in \%). EFPR is computed by discounting misidentifications that do not impact the trace.}
\label{table:function-accuracy}

{\renewcommand{\arraystretch}{1.2}
\begin{threeparttable}
\begin{tabular}{c c c c | c c c c}
\hline
{\bf ID} & {\bf \# Fns } & {\bf TPR} & {\bf EFPR} & {\bf ID} & {\bf \# Fns } & {\bf TPR} & {\bf EFPR}\\
\hline
0a1 & 324 & 100 &  0.29 & 1af & 1674 & 95.52 & 0.99 \\
1d7 & 951 & \textit{93.27} & 0.97 & 3b1 & 204 & 99.02 &  0 \\
443 & 598 & 100 & 0 & 4d7 & 1563 & 95.71 & 1.17 \\
589 & 1486 & 97.51 & 0.68 & 5d3 & 398 & 99.50 & 0.73 \\
646 & 166 & 98.80 &  0 & 67e & 2138 & 99.16 &  0.05 \\
681 & 1961 & 97.86 & 0.56 & 6ac & 265 & 98.11 &  0.37 \\
70b & 115 & 95.65 &  0 & 7e8 & 1529 & 97.58 &  0.66 \\
928 & 520 & 95.38 & 0.93 & 938 & 2764 & 99.57 & 0.74 \\
989 & 762 & 95.80 & 9.7* & ade & 1951 & 99.33 & 0.89 \\
bad & 839 & \textit{92.25} &  0.79 & be7 & 2035 & 99.71 &  0.39 \\
cb5 & 92 & \textit{94.57} &  1.11 & cc8 & 1582 & \textit{94.82} & 0.71 \\
dd9 & 801 & 96.63 & 6.9* & e2a & 495 & 95.15 & 0.39 \\
e2d & 698 & 96.42 &  0.35 & f2b & 1926 & 99.79 &  0.65 \\
f37 & 1585 & 95.21 & 1.16 & fe9 & 1007 & 99.40 &  0.1  \\

\arrayrulecolor{lightgray}\hline

\arrayrulecolor{black}\hline
\end{tabular}
ID = First three characters of SHA256 of binary. \#Fns = Number of functions.

\end{threeparttable}
}
\end{table}

\subsection{Accuracy of Function Block Identification}
We evaluate the accuracy of \texttt{argXtract}'s function block identification (\S\ref{subsection:id-function-blocks}) by identifying function blocks, i.e., function start addresses, for the 28 stripped binaries within our test set and comparing them against the actual functions from the unstripped versions.

We present our results in Table~\ref{table:function-accuracy}. Considering the True Positive Rate (TPR), the table shows that for all but four test binaries, more than 95\% of functions are correctly identified. Manual analysis of the four cases with a TPR lower than 95\% (italicised within table) showed that the functions that were not identified were of unusual structure, e.g., functions accessed via direct conditional branches, or ``functions'' containing only a \texttt{bx} \texttt{lr} instruction. These are likely to be fragments of other functions or shared functions. 

When manually analysing the false positives (FPs) obtained by \texttt{argXtract}, we found that the vast majority of misidentified functions were either where unannotated data had been identified as the start of function blocks, or where a logical function start \textit{can} be assumed, e.g., blocks of alternating \texttt{ldr} instructions and data bytes causing each \texttt{ldr} to be considered as the start of a new function. In the former case, these particular ``functions'' will never be called during the register tracing phase. In the latter, the functions \textit{are} directly addressed as if they are individual functions. Therefore, such FPs will not affect the trace. We thus consider an ``Effective FPR'' to denote the false positives \textit{excluding} such instances. All but 2 binaries had EFPRs of $<1.5\%$. The two test binaries that resulted in an EFPR greater than 1.5\% (marked with * in the table) were both compiled by IAR, which is the only compiler we have observed that uses \texttt{bl} instructions to branch and link \textit{within} a function. This accounts for the higher EFPR for these two binaries. While this will not impact the actual branching functionality, it will influence the call depth calculations, which in turn \textit{could} impact tracing.

\subsection{Correctness of Results}
\label{subsection:correctness}
For correctness checks, we perform tests using binaries generated with known configurations, as well as verification using a real-world binary and associated device.

We use a subset of ten binaries from our test set, targeting Nordic and STMicroelectronics chipsets, compiled using GCC, Keil and IAR, and implementing ANT and BLE. 
For the ANT binaries, we define different channel settings and encryption keys. For Nordic BLE binaries, we define 3 BLE services with very specific configurations as follows:
\begin{itemize}[noitemsep]
    \item Heart Rate Service: must include the Heart Rate Measurement and Body Sensor Location characteristics.
    \item Device Information Service: must include the Manufacturer Name String characteristic and \textit{no other characteristics} (Nordic defines all possible characteristics associated with the Device Information Service within its BLE stack. However, an accurate trace should only identify those characteristics that we have explicitly included within our code). 
    \item Custom service: must include our custom characteristic.
\end{itemize}
Each characteristic also has specific permissions.
It is only if all these configurations are identified exactly that an output is taken to be correct.  
For STMicroelectronics BLE binaries, we define different advertising addresses and privacy configurations. Note that the configurations in each case depended on the options made available by the respective vendors.

In our experiments, we found that all of the conditions were satisfied for all test binaries within our control set, i.e., the configurations were extracted exactly as expected. 

We additionally verified \texttt{argXtract} against the Goji Go Activity Tracker, whose firmware we had extracted from its companion mobile application. The tracker had two SIG services, as well as the Nordic DFU service and a developer-defined service. Comparing the results obtained using \texttt{argXtract} with those we obtained from manual analysis (via a combination of device interaction using the nRF Connect app and profiling using our open-source tool \texttt{ATT-Profiler}~\cite{Sivakumaran:2018:Att-Profiler}, we found that \texttt{argXtract} accurately extracted the configuration of the device.

\section{Case Study: BLE Security and Privacy (Nordic)}
\label{section:nordic}
In this section we present a case study for the identification of BLE configuration vulnerabilities in binaries that target Nordic Semiconductor chipsets. For its BLE offerings, Nordic provides a BLE stack to which configuration requests are issued using supervisor calls.

\paragraph{Building the firmware dataset}
As we have observed previously, IoT peripherals typically interface with one or more mobile applications. 
Many of these mobile applications implement a firmware upgrade and/or factory reset procedure. 
The firmware used for this purpose is either included within the mobile application itself or is downloaded from a server. The firmware for Nordic chipsets is identifiable due to its specific structure and included files.

We programmatically extracted 243 unique\footnote{Determined by the SHA256 computed over the file bytes.} Nordic BLE binaries from a large dataset of BLE-enabled Android APKs, obtained from AndroZoo~\cite{Allix:2016:Androzoo} and Google Play.
We check for the possibility of cloned firmware or different versions of firmware from a single developer, which can result in the same set of characteristics in different files. We use \texttt{ssdeep} for this purpose, with a threshold of 70\%, to account for the fact that a lot of the Nordic baseline code will likely be the same across files. The output showed that 7 clusters were present within our dataset, with an average of 3 files per cluster. We account for these, where relevant, when presenting our results.

\paragraph{Execution environment}
We executed \texttt{argXtract} on a VM running Ubuntu 18.04.3 LTS with 64 GB RAM and 10 processor cores. Taking RAM usage into consideration, 8 parallel processes were used. 

\paragraph{Section outline}
The remainder of this section describes our findings. We first review the protection (or lack thereof) applied to BLE data across the binaries for the link and application layers (\S\ref{subsection:ble-char-security}). We then analyse instances of weakened pairing due to the use of fixed passkeys (\S\ref{subsection:fixed-passkeys}). Finally, we examine privacy concerns identified for our dataset due to two reasons: static addresses (\S\ref{subsection:user-tracking}) and device/manufacturer names (\S\ref{subsection:device-name-privacy}). Each subsection describes the supervisor call that was targeted, mentions the obtained results, and discusses the security or privacy implications of the results.

\subsection{Security of BLE Data}
\label{subsection:ble-char-security}
BLE data is stored in discrete structures known as attributes. \textit{Characteristics} are a type of attribute that hold the data of interest, e.g., heart rate readings. 
Multiple characteristics are grouped into a \textit{service}, which is also a type of attribute. A third type of attribute, \textit{descriptors}, describes a characteristic value. 
An attribute has three types of permissions applied to it: (i) \textit{Access permissions} control whether the attribute can be read and/or written; (ii) \textit{Authentication permissions} specify the required level of protection for the link between the two devices before the attribute can be accessed; (iii) \textit{Authorisation permissions} require developer-specific checks and can be used to implement app-layer security.

When link-layer protection is required (i.e., via authentication permissions), three security \textit{modes} can be applied. We discuss only Mode 1 in this study as we have not observed Modes 2 and 3 in real-world devices. 
The Bluetooth specification defines four levels of protection for Mode 1:
\textit{Level 1} - No security; 
\textit{Level 2} - Unauthenticated pairing with encryption, i.e., encryption with no requirement for Man-in-the-Middle (MitM) protection. This can be achieved using the \textit{Just Works} pairing model, which uses an all-zero key as an input to the key derivation algorithm and requires no user interaction; 
\textit{Level 3} - Authenticated pairing with encryption, i.e., encryption with MitM protection. This requires either the \textit{Passkey Entry} or \textit{Numeric Comparison} pairing models, both of which require user interaction; 
\textit{Level 4} - Authenticated LE Secure Connections pairing with encryption using a 128-bit key. 

Services are freely readable but not writable. Characteristics and specific descriptors can have authentication and authorisation permissions. Characteristics also have certain \textit{properties}, to determine how their data can be accessed. For example, a characteristic value can be \textit{read}. It can also be obtained via \textit{notifications} or \textit{indications}, by writing to a descriptor called the Client Characteristic Configuration Descriptor (CCCD), whereby the BLE peripheral informs the connected device of changes in the characteristic value. While the outcome is somewhat similar (i.e., the connected device obtains the characteristic value), the security requirements for the two mechanisms are different. A read request requires that the connected device satisfy the read permissions specified for the characteristic itself, while subscribing to notifications requires that the connected device satisfy the \textit{write} permissions of the characteristic's CCCD.

\paragraph{Extracting characteristic security configurations}
The security requirements for BLE characteristics in Nordic devices are defined when the characteristics are declared. This is achieved using the \texttt{sd\_ble\_gatts\_characteristic\_add} call. To tie each characteristic to a service such that it can be uniquely identified, we also analyse the \texttt{sd\_ble\_gatts\_service\_add} call.

Executing \texttt{argXtract} against our dataset with a maximum execution time of 1.5 hours per trace returned 199 valid output files. Analysis was completed in less than 2 minutes for \~50\% the binaries. 25\% of the binaries required 2-10 minutes, while the remaining required more than 10 minutes. 
The 199 binaries contained 6 of the previously mentioned clusters. Manual examination showed that the corresponding output files within each cluster had the same service configurations. We therefore consider only 188 unique outputs.

\subsubsection{Insufficient Protection for BLE Data}
In this section, we discuss the protection applied to BLE data for the binaries in our dataset. Because BLE characteristics can either be defined by the Bluetooth SIG, with SIG-specified security configurations, or defined by the device developer with developer-specified security, we analyse the two instances separately.

{\textit{(i) Protection for SIG-defined BLE data:}}
 \texttt{argXtract} extracted SIG-defined characteristics from 103 binaries. 
We compared the obtained security configurations with the expected values specified by the SIG and found that in all cases, the devices do follow the SIG specifications, in that most characteristics have no security applied to them. 

The results also revealed an interesting observation for the SIG-defined characteristics that \textit{do} have security requirements. 
In many such cases, the SIG specifies a choice of protection levels, normally Mode~1 Level~2 \textit{or} Mode~1 Level~3. 
These can be achieved using \textit{Just Works} or \textit{Passkey Entry} pairing, respectively. 
While both \textit{Just Works} and \textit{Passkey Entry} are known to be vulnerable to passive eavesdropping attacks~\cite{Bluetoothsig:2019:Specification2}, \textit{Passkey Entry} should be the choice for greater security, as it provides some MitM protection. 
However, our results show that device developers have invariably opted for the \textit{lower} security level, i.e., Mode~1 Level~2. 
This may be due to insufficient IO capabilities on the devices precluding the use of \textit{Passkey Entry}, although we have previously observed real-world devices using \textit{Just Works} even when sufficient IO capabilities \textit{were} available. 
We reiterate that, even if the BLE device \textit{does} have sufficient IO capabilities, so long as Mode~1 Level~2 is specified, an attacker can manipulate the pairing process such that \textit{Just Works} is used.

\begin{table}[!t]
\centering
\footnotesize
\caption{Binaries with protection applied to developer-defined data.}
\label{table:vendor-chars}

{\renewcommand{\arraystretch}{1.2}
\begin{threeparttable}
\begin{tabular}{p{0.55cm} p{6.1cm} c}
\hline
& & {\bf \#Bins}  \\
\hline
\textbf{\textit{Reads}} 
& Binaries with developer-defined readable characteristics & 167 \\
& Binaries with Mode1-Level2 link-layer protection for developer-defined readable characteristics & 5 \\
& Binaries with Mode1-Level3 link-layer protection for developer-defined readable characteristics & 0 \\
& Binaries with application-layer security for developer-defined readable characteristics & 7 \\
\hline
\textbf{\textit{Writes}} 
& Binaries with developer-defined writable characteristics & 169 \\
& Binaries with Mode1-Level2 link-layer protection for developer-defined writable characteristics & 4 \\
& Binaries with Mode1-Level3 link-layer protection for developer-defined writable characteristics & 0 \\
& Binaries with application-layer security for developer-defined writable characteristics & 69* \\
\hline
\end{tabular}
* 24 excluding Nordic DFU control point, from Nordic DFU library.

\end{threeparttable}
}
\end{table}

{\textit{(ii) Protection for developer-defined BLE data:}}
\label{subsubsection:results-vendor-data} \texttt{argXtract} extracted at least one developer-defined characteristic from 170 binaries. Table~\ref{table:vendor-chars} summarises the link layer and application layer protection applied to the developer-defined characteristics, broken down into readable (including notifications/indications) and writable characteristics.

From the table, we conclude that protection for reads is virtually non-existent at the link-layer, with only five firmware binaries specifying Mode~1 Level~2 authentication requirements. Similarly, authorisation requirements are also not prevalent among readable characteristics, with only seven binaries specifying protection at higher layers. 

In terms of link-layer protection, writable characteristics fare similarly to readable characteristics, with only four binaries specifying Mode~1 Level~2 authentication requirements. App-layer protection is slightly better for writable characteristics, but a significant proportion of binaries apply no protection at all to their writable characteristics (apart from those provided by Nordic itself, for its firmware upgrade procedure). 

\paragraph{Security implications}
The security of a BLE device is strongly associated with the authentication and authorisation permissions applied to its data.
\textbf{\textit{Having freely accessible BLE characteristics means any user in the vicinity of the BLE peripheral will be able to read and write the data}}, subject to the characteristic being readable/writable. For that matter, even if the characteristic is protected by \textit{Just Works} pairing, an attacker in the vicinity can pair with the device and access its data. We verified this with a fitness tracker, from which we were able to access characteristic values without pairing. 
Further, even if strong link-layer protection was present, \textbf{\textit{absence of application-layer protection makes the data vulnerable to access by unauthorised apps}}~\cite{Sivakumaran:2019:Crossapp}. We verified this with a custom Android app and an emulated BLE device.

\vspace{2mm}

\textit{Implications for readable data within the dataset}: Among the binaries that had no link-layer or application-layer protection for readable characteristics, we found numerous fitness tracking devices, as well as healthcare devices, all of which potentially store detailed information regarding a user's activity or health. No protection or only \textit{Just Works} protection means that this personal and sensitive data is vulnerable to unauthorised access, via local \textit{and} remote attacks.

\vspace{2mm}

\textit{Implications for writable data within the dataset}: Within the binaries that had writable characteristics, we found one that contained the SIG-defined Human Interface Device (HID) service. This only had Mode~1 Level~2 link-layer protection applied to its characteristics. Again, this security requirement can be satisfied by \textit{Just Works} pairing, which means that an attacker could transmit unsolicited messages to the HID device, and also read and modify the keyboard characters that are transmitted between the HID device and host via a MitM attack. This has been demonstrated in~\cite{klostermeier:2018:security}. We have informed the HID device developer of this vulnerability, but have not yet received a response.

The vast majority of devices applied no protection to \textit{any} of their writable characteristics. Among these were smart switches, medical respiratory devices (nebulisers) and ECG monitoring devices. Writing random bytes to characteristics on such devices could cause the devices to function improperly or cease to function entirely. If the behaviour of the device corresponding to the written values is known to an attacker, then they can write carefully chosen values in order to modify the expected behaviour, possibly with harmful consequences.

\subsubsection{\textbf{Different Permissions for Read vs. Notify}}
\label{subsubsection:results-read-notify}
As mentioned previously, the value held within a characteristic can be accessed in different ways, either using a direct read request or via notifications/indications, and even though the mechanisms of access differ, the ultimate outcome is similar. We observed that one binary within our results contained characteristics that had both \texttt{read} \textit{and} \texttt{notify} properties, but with different security properties set for the two types of access. Mode~1 Level~2 security was required to be able to \textit{read} the characteristics' values, while the values could be freely accessed via \textit{notifications} (their CCCDs were writable without the need for any pairing or higher-layer protection). 

\paragraph{Security implications}
Different security levels for different value acquisition methods implies that the data can always be accessed using the less secure mechanism. In addition, there may be a false sense of security, as the protection will be assumed to be higher than it actually is. This finding shows that \textbf{\textit{developers may unintentionally leave ``gaps'' in security}}, particularly when incorporating different functionalities. We have informed the BLE device developer about this issue, but have not yet received a response.

\subsection{Use of Fixed Passkeys}
\label{subsection:fixed-passkeys}
As mentioned in~\S\ref{subsection:ble-char-security}, \textit{Passkey Entry} is a BLE pairing model which provides MitM protection by requiring that a user manually key in a passkey that is displayed on the BLE peripheral. However, some developers choose to program a fixed passkey into the peripheral. This might be because many BLE peripherals don't have IO capabilities (i.e., keypad or display), but could also originate from bad practices when programming devices that \textit{do} have these capabilities.

\paragraph{Identifying fixed passkeys}
The \texttt{sd\_ble\_opt\_set} supervisor call enables setting multiple options on Nordic BLE devices, where each option is identified using an opt\_id. The opt\_id with value 34 denotes setting a fixed passkey. \texttt{argXtract} identified a binary within the dataset that contained a fixed passkey of \texttt{0x303030303030}, i.e. ``000000''. 

\paragraph{Security implications}
Fixed passkeys undermine the security of the \textit{Passkey Entry} model, particularly if the same passkey is used for all devices of a certain brand. In such a scenario, an attacker would only need to know the passkey for one device in order to be able to covertly connect to any device of the same brand. This effectively \textbf{\textit{removes the MitM protection afforded by the \textit{Passkey Entry} model}}. With the binary we identified, a fixed passkey of ``000000'' equates \textit{Passkey Entry} to the \textit{Just Works} model (since \textit{Just Works} uses an all-zero key).

\subsection{User Tracking due to Fixed Addresses}
\label{subsection:user-tracking}
BLE peripherals periodically transmit messages on advertising channels in order to enable incoming connections. These messages contain the peripheral's hardware address and could be used to track the device. To overcome this, the BLE specification defines \textit{resolvable private addresses}, which enable a device to change its advertising address while still allowing for reconnections from bonded peers. 

There are in fact four types of addresses that can be used with BLE: Public, Random Static, Private Resolvable, and Private Non-resolvable. Public addresses do not change during the lifetime of a device; Random Static addresses do not change during a single power cycle and may not change for the lifetime of the device; Private Resolvable addresses change periodically in such a way as to enable reconnections by a bonded peer; Private Non-resolvable addresses change periodically, but do not allow for reconnections.

\begin{table}[!t]
\centering
\footnotesize
\caption{Address types used in BLE peripherals.}
\label{table:addresses}

{\renewcommand{\arraystretch}{1.2}
\begin{threeparttable}
\begin{tabular}{l c l c}
\hline
{\bf Address Type} & {\bf \#Binaries} & {\bf Address Type} & {\bf \#Binaries} \\
\hline
Public & 29 & Private nonresolvable & 1 \\
Random static & 208 & Private resolvable & 1 \\

Unknown & 4 \\
\hline
\end{tabular}

\end{threeparttable}
}
\end{table}

\paragraph{Extracting advertising address type}
By default, Nordic uses a random static address for each chipset, which is set at the time of manufacture and does not change for the lifetime of the device. However, developers are able to modify this behaviour to choose a different address type via the \texttt{sd\_ble\_gap\_address\_set} (in older versions of the stack) and the \texttt{sd\_ble\_gap\_addr\_set} and \texttt{sd\_ble\_gap\_privacy\_set} calls. 

\texttt{argXtract} extracted the arguments to these supervisor calls (where present) for all binaries within the dataset. 
35 out of the 243 firmware files included one of the \texttt{svc} numbers for performing address type selection/setting, which meant that the remaining 208 files used the default setting of a random static address, set at the time of manufacture and unchanging throughout the lifetime of the device. 

Table~\ref{table:addresses} depicts a breakdown of the address types used within the BLE peripherals in our dataset. Out of the 243 binaries in our dataset, only a single binary used resolvable private addresses. Combining this with information regarding the device name, we found that the binary was related to a personal protection device. One binary within the dataset used non-resolvable addresses, which means it will not be vulnerable to tracking but will also not be able to form bonds with other devices. We could not deduce its functionality from its device name. We found that overall, the results indicated that at least 95\% of the BLE binaries use static (random or public) addresses.

\paragraph{Privacy implications} 
Because BLE peripherals tend to advertise constantly when not in a connection, \textbf{\textit{the use of public or random static addresses in advertising messages opens the BLE device, and by extension (depending on the device) its owner, to tracking}}. 
Further, even private resolvable addresses may be vulnerable if the associated address resolving key is not sufficiently protected.
In crowded locations such as shopping centres, repeated visits by a user can be covertly tracked simply by monitoring BLE advertisements and logging the device addresses. This has been previously demonstrated in~\cite{das:2016:blePrivacy}. It has also been shown to be feasible to set up a botnet to track users across a range of locations~\cite{Issoufaly:2017:bleb}.

These attacks are particularly relevant in the case of devices such as wearables, which are generally always on the user's person. We found that \textbf{\textit{all of the wearable binaries within our dataset used public or fixed random static addresses}}. 

\subsection{Manufacturer/Device Names and Privacy}
\label{subsection:device-name-privacy}
BLE advertising messages usually contain the peripheral's name, which is often used by users to identify a device from (potentially) a number of other BLE devices that are also advertising in the vicinity. Peripherals may also include a Manufacturer Name String, normally obtained via a scan request. These advertising messages require no authentication in order to be read.

\paragraph{Extracting device and manufacturer names}
A Nordic BLE device's name is set programmatically using the \texttt{sd\_ble\_gap\_device\_name\_set} supervisor call, while the Manufacturer Name String is included within the Device Information Service (obtained in \S\ref{subsection:ble-char-security}). \texttt{argXtract} extracted non-null values for at least one of device or manufacturer name from 156 binaries. 

An analysis of the device names revealed that our dataset contained firmware from a variety of BLE devices, including wearable fitness trackers, beacons, electric switch controls, parking aids, security devices, personal protection devices, medical equipment and behavioral monitoring devices. 

\paragraph{Privacy implications}
Device names can reveal a lot about the nature of the device. This is particularly concerning when the device is related to a user's health, or is of an otherwise private nature. Because no active connections are required to read advertising data, an attacker would simply need to monitor the BLE advertising channels and perhaps send a scan request for additional information. By continuously scanning BLE advertisements, extracting the device and manufacturer name, and combining this information with the Received Signal Strength Indicator (RSSI), along with user observation, an attacker may be able to determine which devices belong to which users in the vicinity. 
This could defeat resolvable private addresses, as an attacker might instead be able to use the device name, along with other advertising data, to track the device~\cite{fawaz:2016:protecting, Celosia:2019:Fingerprinting, Becker:2019:Tracking}.
Further, if a particular device has known issues (such as those identified in this thesis), then the attacker can take advantage of the device name to identify exploitable devices.

\section{Applicability Case Studies}
\label{section:applicability}
\texttt{argXtract} is not confined to a single technology or vendor. To demonstrate its applicability for other technologies and vendors, we present here two studies targeting BLE binaries for STMicroelectronics' BlueNRG chipsets and ANT binaries for Nordic chipsets.

\subsection{BLE Security and Privacy (BlueNRG)}
\label{section:bluenrg}
In this section we present a case study for the identification of BLE configuration vulnerabilities in firmware that targets STMicroelectronics chipsets, specifically the BlueNRG family of processors. Configurations for BlueNRG are performed via function calls. We therefore use the function pattern matching feature of \texttt{argXtract} (\S\ref{subsubsection:pattern-matching}) to determine the location of the relevant function within the disassembly.

We manually analysed 500 real-world \texttt{.bin} files extracted from APKs for ARM vector tables. Two such binaries were found to be STMicroelectronics BlueNRG binaries. \texttt{argXtract} identified that both had an application code base of \texttt{0x10051000}, which corresponds to BlueNRG-1 v2.1+~\cite{STM:2018:AN4869}.

\subsubsection{BLE Address Privacy}
In this section, we describe our tests against the BlueNRG binaries to identify the use of resolvable private addresses, which are enabled by the \texttt{aci\_gap\_init} function.

\paragraph{Extracting address configurations}
BlueNRG provides the \texttt{aci\_hal\_write\_config\_data} function for configuring public addresses~\cite{STM:2019:PM0257}. This function takes three arguments: an offset, a length and a pointer to a byte array. In the case of address configuration, the offset is expected to be 0, the length is 6 and the byte array contains the address. This can be used to validate obtained outputs. Internally, this function calls \texttt{HAL\_Write\_ConfigData} with the same arguments. The output memory contains the configured address at a binary-specific location. 
Privacy is configured separately via the \texttt{aci\_gap\_init} function. This takes three inputs: the BLE role of the device, an integer indicating whether privacy is enabled (i.e., whether resolvable addresses are used), and the length of the device name. The function performs several tasks, most of which require runtime information. However, we exploit the fact that the function adds the GAP service to the database and specify test sets that look for the service structure within the output memory contents.

Executing \texttt{argXtract} against the real-world binaries revealed that one contained a public address derived from the BlueNRG example address. This along with the binary's name led us to conclude that the binary was for demonstration purposes. The second binary was a BLE-enabled cyclist safety aid. It did not have privacy enabled.

\paragraph{Privacy implications}
A cyclist safety aid is likely to be about the user's person whenever they are cycling. A fixed address emanating from the device at all times enables the user to be tracked over time, as discussed in \S\ref{subsection:user-tracking}.

\subsubsection{BLE Pairing Security}
With BlueNRG binaries, if a BLE characteristic has authentication requirements, then specific configurations must be performed to enable pairing. We exploit two pairing-related functions in our tests. 
We additionally check for authorisation requirements.

\paragraph{Extracting pairing configurations}
BlueNRG requires two function calls in order to enable BLE security: the first is the \texttt{aci\_gap\_set\_io\_capability} to set the device's input-output capability, and the second is the \texttt{aci\_gap\_set\_authentication\_requirement} to set the pairing requirements (such as bonding, MitM protection, etc)~\cite{STM:2019:PM0257}. Authorisation permissions are set using \texttt{aci\_gap\_set\_authorization\_requirement}.
Focusing on the cyclist safety aid, \texttt{argXtract} found that the binary had no calls to \texttt{aci\_gap\_set\_io\_capability}, nor did it have any calls to \texttt{aci\_gap\_set\_authorization\_requirement}. This means that BLE security was not enabled. 

\paragraph{Security implications}
The fact that no security was present on the cyclist safety aid means that an attacker could connect to the device and send commands to it without the need for any authentication, which could have serious consequences for the cyclist's safety. We have informed the safety aid developer regarding the identified issues, but have not received a response as yet.

\subsection{ANT Security (Nordic)}
\label{section:ant}
ANT is a wireless communication technology designed by ANT Wireless, a division of Garmin Canada. It is intended for personal area networks and is used predominantly in fitness and sports sensor applications. We focus on the ANT stack provided by Nordic Semiconductor, which uses supervisor calls for configuration.

To acquire Nordic ANT binaries, we follow the same procedure as for Nordic BLE, but focus on a different set of \texttt{svc} numbers. We obtained 9 ANT binaries from APKs.

\subsubsection{ANT Channel Security} ANT communications are channel-based, with a channel connecting two or more nodes together. Some ANT devices can have multiple channels. To secure the channels at the network layer, ANT supports 8-byte network keys and 128-bit AES encryption~\cite{ANT:2020:ANTSecurity}.
 
\paragraph{Extracting channel security configurations:}
ANT is enabled using the \texttt{sd\_ant\_enable} supervisor call. Within this call, the number of required channels and the number (out of those created) that should be encrypted are specified. 

\texttt{argXtract} extracted channel configuration parameters from 9 real-world ANT binaries, corresponding to 7 indoor exercise bikes, an analytical bike light (i.e., a bike light with additional sensors), 
and a heart rate monitor. Three binaries defined a single ANT channel, four defined 2 channels and two defined 4 channels. \textbf{\textit{None of the binaries specified encryption for any of their ANT channels}}.

\paragraph{Security implications:}
As with the findings discussed in \S\ref{subsection:ble-char-security} for BLE, in ANT too data will be vulnerable to unauthorised access if channel security is not enabled. One of the binaries that was tested was a heart rate monitor, which means that a user's heart rate measurements - which are an indicator of the user's health - are vulnerable.

\section{Limitations and Future Work}
\label{section:discussion}
\paragraph{Edge cases}
\texttt{argXtract} is able to analyse most Cortex-M binaries. However, there are edge cases where the \verb+.text+ segment is split into sub-sections, with different address offsets for each sub-section, where \texttt{argXtract} is unable to obtain individual code bases and accurate function estimates. This improvement is left as future work.

\paragraph{Function pattern matching}
As mentioned in \S\ref{subsection:coi-identification}, the function pattern matching performed by \texttt{argXtract} uses manually-defined test sets, when function outputs or artefacts are distinguishable. 
The function pattern matching process can take several hours when a binary contains a large number of functions.
An ideal alternative would be \textit{automated} function pattern matching, \textit{without} executing function code. 
The most popular method to achieve this at present is via machine learning techniques. 
However, this requires a sufficiently large annotated training set for each function of interest, which is not yet available for the types of vendor-specific configuration functions that are of interest here.   

\paragraph{Register tracing}
Assigning a value of 0 to register \texttt{r0} when a branch is not followed due to exceeding the $max\_call\_depth$ (as described in \S\ref{subsection:arg-matching}) can give rise to inaccurate results if the value is actually supposed to be non-zero.
We have not encountered such errors for our use cases, because the data used as arguments to our COIs did not go through several nested branches of processing.
If it is known that COI arguments are likely to be highly-processed, then increasing the $max\_call\_depth$ can reduce the likelihood of errors at the expense of (much) longer trace times.

\paragraph{Binary patching}
\texttt{argXtract} at present is able to extract arguments to security-relevant configuration APIs.
A natural extension would be to perform patching of the binary if vulnerable configurations \textit{are} identified.
This may be fairly straightforward in cases where the relevant API call is present within the binary and the configuration only requires modification of simple input structures. 
However, composite input structures would require complex handling. 
In addition, vendor-specific considerations regarding CRC and other correctness checks may also have to be accounted for.

\section{Related Work}
\label{section:related-work}
Various aspects regarding the analysis of firmware binaries and configuration security have been explored in previous studies. While it may seem like most aspects of firmware analysis have already been covered by previous studies, most such studies have focused on Linux-based systems~\cite{qasem:2021:automatic}. We observe that the analysis of stripped binaries targeting non-traditional operating systems and the ARM Thumb instruction set, which is increasingly favoured by IoT peripherals and which is the focus of our analysis, has still not been explored sufficiently. 

\textit{Analysis of stripped binaries:}
The analysis of stripped binaries, particularly function block identification, has been the subject of widespread study. Control flow analysis has been used in~\cite{prasad:2003:binary, harris:2005:strippedbinary, ravipati:2007:dyninst, qiao:2017:function, andriesse:2017:compiler} to determine functions in PE, ELF, COFF and XCOFF binaries, and a QEMU+LLVM approach for function boundary identification was presented in~\cite{di:2017:rev}. These approaches may not be suited to ARM IoT analysis due to errors introduced by inline data and compiler-introduced constructs such as Thumb switch-case conditions. 
Machine learning (ML) has also been proposed for identifying function entry points~\cite{rosenblum:2008:learning, bao:2014:byteweight, shin:2015:recognizing}, but this approach requires a sufficiently large labelled training set, which is currently not available for IoT peripheral binaries. 
A semantics-based approach was used in Jima~\cite{alves:2019:jima} for ELF x86/x86-64, which employs techniques for computing jump tables that are similar to those used in \texttt{argXtract} for computing table branch addresses.

\textit{Function matching and labelling:}
Previous works have employed various strategies to achieve function pattern matching or similarity computations (primarily for non-ARM binaries). 
One approach for function pattern matching is to compute statistical similarities between instruction sequences of functions~\cite{myles:2005:kgram, lee:2013:function}, but this may suffer poor performance due to compiler-introduced variations and optimisations~\cite{bourquin:2013:binslayer}. 
Dynamic similarity testing via function execution was employed in~\cite{egele:2014:blanket}. While this is in some ways similar to our approach, \texttt{argXtract} looks for functional \textit{equivalence} based on \textit{known} function behaviour, while~\cite{egele:2014:blanket} considers function \textit{similarity} based on \textit{random} executions. 
Most current approaches favour ML techniques~\cite{shin:2015:recognizing, xu:2017:neural, massarelli:2019:safe, patrick:2020:probabilistic} but, as mentioned previously, this requires sufficiently large training sets.

\textit{Security analysis of IoT firmware:}
A number of previous works have described techniques for assessing the security of IoT devices via static or dynamic firmware analysis. 
Large-scale security analyses of embedded firmware files, predominantly Linux and VxWorks-based, were presented in~\cite{costin:2014:embeddedfirmware, costin:2016:automated}. FIE~\cite{davidson:2013:fie}, built from the KLEE symbolic execution engine, identifies vulnerabilities in embedded MSP430 firmware. Firmalice~\cite{shoshitaishvili:2015:firmalice} detects authentication bypass vulnerabilities within the firmware of Linux and VxWorks-based binaries. 
FirmFuzz~\cite{srivastava:2019:firmfuzz} specifically targets IoT firmware and is intended for security analysis. It uses QEMU and targets unstripped Linux-based binaries. 
These works analyse binaries that target at least pared-down versions of fully-fledged operating systems. They would not be suitable for analysing stripped firmware of embedded devices that do \textit{not} have a proper OS. 
\texttt{InternalBlue}~\cite{Mantz:2019:internalblue} enables testing and patching of Broadcom Bluetooth firmware, while \texttt{LightBlue}~\cite{wu:2021:lightblue} analyses and performs debloating of unneeded Bluetooth HCI commands within firmware to reduce the potential attack surface. The randomness of RNGs used in Bluetooth chipsets was measured via firmware analysis in~\cite{tillmanns:2020:firmware}.

\textit{Analysis of BLE binaries:}
On the BLE front, previous works (including our own) have explored the security and privacy configurations and behaviour of BLE peripherals by analysing devices~\cite{das:2016:blePrivacy, fawaz:2016:protecting,Antonioli:2020:knoble, wang:2020:bluedoor,classen:2018:anatomyfitbit, Hilts:2016:EveryStep, Cyr:2014:Security}, and mobile apps~\cite{Sivakumaran:2019:Crossapp, zuo:2019:uuid}. However, device analysis is expensive and may not be generalisable for bulk analysis, while mobile applications generally don't provide insights about low-level pairing mechanisms.

Independently to us, Wen, et al.~\cite{wen:2020:firmxray} developed a tool named \texttt{FirmXRay} that identifies BLE link layer configuration vulnerabilities by targeting supervisor calls on Nordic and ICalls on Texas Instruments BLE binaries. 
While \texttt{FirmXRay} is geared towards BLE vulnerabilities, our work is capable of handling generic analysis of any technology that targets ARM Cortex-M binaries. 
Further, \texttt{FirmXRay} only handles supervisor calls and ICalls, whereas \texttt{argXtract} perform function pattern matching to identify \textit{any} function (provided the requisite artefacts can be identified within memory/registers). The template-based approach used in our tool also enables easy addition of new test functions.
Within the BLE analysis, Wen, et al.~\cite{wen:2020:firmxray} have confined the discussion to link layer vulnerabilities, while we discuss app-layer issues as well.

\section{Resources}
\label{section:resources}
\texttt{argXtract} is available as an open-source tool at

\smallskip

\url{https://github.com/projectbtle/argXtract}

\smallskip

Function pattern files and argument definition files for all use cases described in this paper are also on the repository.

\section{Acknowledgements}
This research has been partially sponsored by the Engineering and Physical Sciences Research Council (EPSRC) and the UK government as part of the Centre for Doctoral Training in Cyber Security at Royal Holloway, University of London (EP/P009301/1).

\section{Conclusion}
\label{section:conclusion}
In this work, we present \texttt{argXtract}, a tool for performing partial-knowledge automated analyses of stripped IoT binaries, to extract configuration information from ARM \mbox{Cortex-M} binaries. \texttt{argXtract} overcomes the challenges inherent to the analysis of stripped binaries and enables bulk processing of IoT firmware files. We use \texttt{argXtract} to extract security-relevant configurations from three datasets: Nordic Bluetooth Low Energy (BLE) binaries, STMicroelectronics BLE binaries and Nordic ANT binaries. Our results reveal widespread lack of protection for data, inconsistent data access controls and serious privacy vulnerabilities. We posit that \texttt{argXtract} paves the way for easier automated security analyses of stripped IoT binaries.

\begingroup
\raggedright
\bibliography{main} 
{\small
\bibliographystyle{IEEEtran}
}
\endgroup
\end{document}